\documentclass[10pt]{iopart}
\usepackage{graphicx}
\usepackage{iopams}
\usepackage{citesort}
\begin{document}

\title{Motion and wake structure of spherical particles}

\author{Christian Veldhuis\dag, Arie Biesheuvel\ddag, Leen van Wijngaarden\dag\ and Detlef Lohse\dag\
\footnote[3]{To whom correspondence should be addressed
(d.lohse@utwente.nl)} }

\address{\dag\ Department of Applied Physics, University of Twente, Enschede, The Netherlands}

\address{\ddag\ Department of Mechanical Engineering, University of Twente, Enschede, The Netherlands}

\begin{abstract}
This paper presents results from a flow visualization study of the
wake structures behind solid spheres rising or falling freely in
liquids under the action of gravity. These show remarkable
differences to the wake structures observed behind spheres held
fixed. The two parameters controlling the rise or fall velocity
(i.e., the Reynolds number) are the density ratio between sphere and
liquid and the Galileo number.

\end{abstract}

\section{Introduction}
This year's cover illustration shows the boundary layer separation
from a light sphere rising in quiescent water at high Reynolds
number. The wake structure behind the sphere is visualized with the
Schlieren technique. The experiment is part of a project to study
wake structures behind falling and rising spherical
particles in a large range of Reynolds numbers.\\
\\
In the past years extensive numerical investigations
\cite{Johnson,Lee,Tomboulides,Dusek,Ploumhans} have established how
the wake of a sphere held {\em fixed} in a uniform flow undergoes a
series of transitions as the Reynolds number $ Re = Ud/\nu $ is
increased. Here $U$ is the free stream velocity, $d$ the diameter of
the sphere, and $\nu$ the kinematic viscosity of the water. It was
found that the wake is axially symmetric up to $Re=212$. Above this
value a planar-symmetric wake is found that consists of two steady
counter-rotating threads. At $Re\approx270$ there is a further
transition and the planar-symmetric flow becomes time-dependent:
opposite-signed streamwise vortices then form a series of loops that
resemble hairpin vortices. As the Reynolds number is further
increased, the flow gradually becomes more irregular and finally
turbulent. The Digital Particle Image Velocimetry (DPIV)
measurements by Br\"{u}cker \cite{Bruecker} and the flow
visualization studies by Schouveiler \& Provansal \cite{Schouveiler}
have confirmed most of these numerical results and have further
elucidated the sequence of transitions.

For {\em freely} moving spheres the Reynolds number is defined by
the measured {\em mean} velocity of rise or fall of the sphere $U_T$
and the corresponding `Reynolds number' becomes $ Re_T =
{\langle U_T \rangle d}/ {\nu}$. The {\em mean} velocity is the
time averaged velocity of the sphere, not including the acceleration
of the sphere from rest. The flow looses its axial symmetry at a
critical Reynolds number which is not significantly affected by the
density ratio $\rho_s/\rho$ \cite{Jenny}: $Re_{cr}$=211.9 for
$\rho_s/\rho \rightarrow \infty$ (i.e. sphere held fixed),
$Re_{cr}=206.3$ for $\rho_s/\rho=0.5$ and $Re_{cr}=205.8$ for
$\rho_s/\rho=0.0$ . This is in good agreement with, for example, the
experimental results on solid spheres \cite{Nakamura}, on
surface-contaminated gas bubbles \cite{Hartunian},  and on wake
visualizations in experiments with drops of tetrachloride and
chlorobenzene falling in water \cite{Magarvey1,Magarvey2,Magarvey3}.
As pointed out by Natarajan \& Acrivos \cite{Natarajan}, these drops
must have behaved effectively as solid spheres due to presence of
surface-active impurities, and these visualizations have therefore
often served as a basis of comparison with numerical studies on
fixed spheres.

What happens for \emph{freely} falling or rising spheres at higher
Reynolds number, which is more common in multiphase flow
applications? How are the wake structures and transitions observed
for the fixed sphere case modified? Is there a (clear) difference in
wake structure between rising and falling spheres? In this paper we
present flow visualizations of the wakes behind freely moving solid
spheres at large Reynolds number ($Re=450-4623$) for which the
density ratio $\rho_s/\rho$ is in the range 0.50 to 2.63.

\section{Experimental details}
\begin{figure}[!t]
\begin{center}
\includegraphics[angle=0, width=0.9\textwidth]{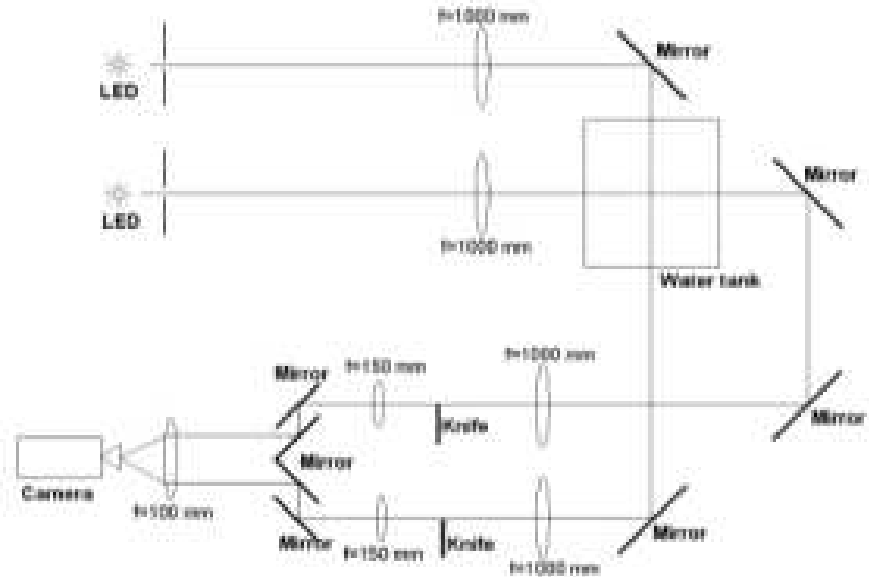}\\
\caption{Top view of the Schlieren set-up used to visualize the
sphere wakes.} \label{SetupFig1}
\end{center}
\end{figure}

The flow visualizations of the sphere wakes were carried out in a
transparent tank (0.15 x 0.15 x 0.5 m$^3$) filled with decalcified
water. Smooth plastic spheres with diameters between 1.5 mm and 10
mm and densities between 500 kg/m$^3$ and 2781 kg/m$^3$ were
released from rest. By means of an optical system consisting of two
LED-lights, pinholes, lenses and mirrors, two perpendicular images
of the particle and its wake were created and recorded at 500
frames/s with a CCD-camera (figure \ref{SetupFig1}). Hence, each
image consists of two perpendicular views of the same sphere. The
images are taken at a position in the transparent tank where the
spheres do not accelerate anymore. The wake was visualized using the
Schlieren technique. To this end a small vertical temperature
gradient in the water was maintained (1 K/cm.). The \emph{mean}
water temperature at the measurement section was 302 K, with
corresponding values of the density ($\langle \rho \rangle $) and
viscosity ($\langle \nu \rangle $) of 996 kg/m$^3$ and
0.802$\cdot$10$^{-6}$ m$^2$/s, respectively. Hence, the Reynolds
number is based on the \emph{mean} viscosity and is defined as $
Re_T = \langle U_T \rangle d/ \langle \nu \rangle $. It turned
out to be difficult to keep a constant temperature gradient.
Therefore the error in the \emph{mean} water temperature at the
measurement section is about 3 K, leading to a relative error in the
viscosity of 10 \%. As opposed to the fixed-sphere problem, the
Reynolds number for freely moving spheres is not an independent
parameter. Following Jenny {\em et al.} \cite{Jenny} we choose as
independent dimensionless variables the ratio $\rho_s/\langle \rho
\rangle$ of the densities and the Galileo number
\begin{equation}
\label{GalileoNumber} G = \frac{{(|(\rho_s/\langle \rho \rangle
-1)|g)}^{1/2}d^{3/2}}{\langle \nu \rangle}.
\end{equation}
Since ${(|(\rho_s/\langle \rho \rangle -1)|g d)}^{1/2}$ can be
considered as a velocity scale, $G$ plays a similar dynamical role
as the free-stream Reynolds number in the case of fixed sphere. The
parameter values for which we made the flow visualizations are
summarized in Table 1.
%\newpage
\begin{table}
\caption{\label{Table1}Parameter values in our visualizations; $d$
in mm, $\rho_s$ in kg/m$^3$. $G$ is defined by equation
(\ref{GalileoNumber}) and $Re_T$ is the \emph{mean} Reynolds number
($ Re_T = \langle U_T \rangle d/ \langle \nu \rangle$).} \lineup
\begin{tabular*}{\textwidth}{@{}l*{15}{@{\extracolsep{0pt plus12pt}}r}}
\br
& Number &  $d$  & $\rho_s$  & $\rho_s/\langle \rho \rangle$ & $G$ &$Re_T$ & Figure\\
\mr
falling & 1 &  3.2  &  1028     & 1.03          & 121 & 205 & Figure \ref{ExtraPicturesFalling}\\
        & 2 &  4.0  &  1058     & 1.06          & 239 & 325 & Figure \ref{ExtraPicturesFalling}\\
        & 3 &  1.5  &  2781     & 2.79          & 304 & 450 & Figure \ref{ExtraPicturesFalling}\\
        & 4 &  6.0  &  1035     & 1.04          & 359 & 546 & Figure \ref{FallingRisingFig2}\\
        & 5 &  6.0  &  1043     & 1.05          & 394 & 608 & Figure \ref{FallingRisingFig2}\\
        & 6 &  4.0  &  2629     & 2.63          &1261 &1970 & Figure \ref{ExtraPicturesFalling}\\
        &   &       &           &               &     &     & \\
rising  & 7 &  3.2  &   965     & 0.97          & 121 & 210 & Figure \ref{ExtraPicturesRising1}\\
        & 8 &  5.0  &   950     & 0.95          & 297 & 450 & Figure \ref{RisingCrossingFig3}\\
        & 9 &  5.0  &   947     & 0.95          & 306 & 475 & Figure \ref{RisingCrossingFig3}\\
        &10 &  4.0  &   873     & 0.88          & 334 & 565 & Figure \ref{ExtraPicturesRising1}\\
        &11 & 10.0  &   988     & 0.99          & 350 & 576 & Figure \ref{RisingKinkFig4}\\
        &12 &  8.0  &   982     & 0.99          & 331 & 602 & Figure \ref{ExtraPicturesRising1}\\
        &13 &  6.0  &   958     & 0.96          & 355 & 647 & Figure \ref{FallingRisingFig2}\\
        &14 &  6.0  &   950     & 0.95          & 390 & 656 & Figure \ref{FallingRisingFig2}\\
        &15 &  6.4  &   925     & 0.93          & 534 & 920 & Figure \ref{ExtraPicturesRising1}\\
        &16 &  6.4  &   864     & 0.87          & 728 &1180 & Figure \ref{ExtraPicturesRising2}\\
        &17 &  7.9  &   925     & 0.93          & 732 &1350 & Figure \ref{ExtraPicturesRising2}\\
        &18 &  6.4  &   650     & 0.65          &1160 &1965 & Figure \ref{ExtraPicturesRising2}\\
        &19 &  9.5  &   500     & 0.50          &2548 &4623 & Figure \ref{ExtraPicturesRising2}\\
\br
\end{tabular*}
\end{table}

\section{Observations}

\begin{figure}[!t]
\unitlength1cm
\begin{minipage}[t]{3.0cm}
\begin{center}
\includegraphics[width=3.0cm]{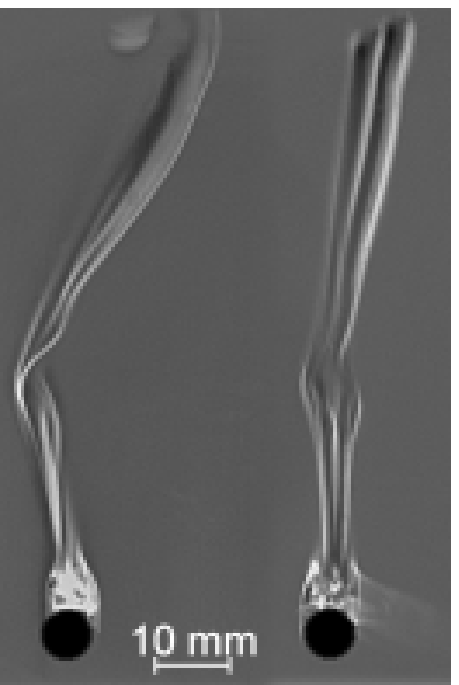}\\
\footnotesize{(a)}
\end{center}
\end{minipage}
\hfill
\begin{minipage}[t]{3.0cm}
\begin{center}
\includegraphics[width=3.0cm]{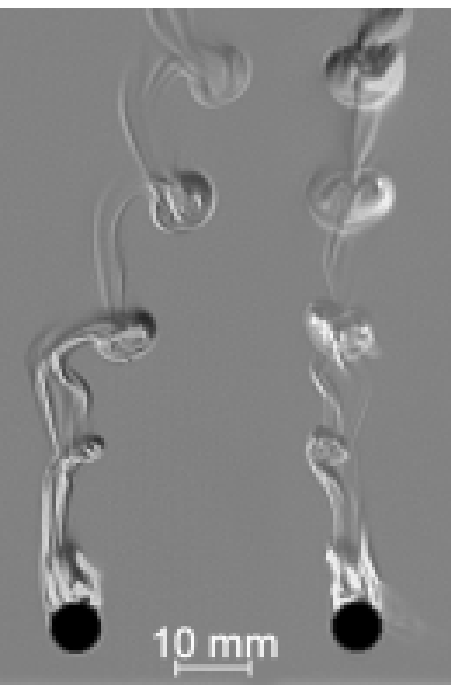}\\
\footnotesize{(b)}
\end{center}
\end{minipage}
\hfill
\begin{minipage}[t]{3.0cm}
\begin{center}
\includegraphics[width=3.0cm]{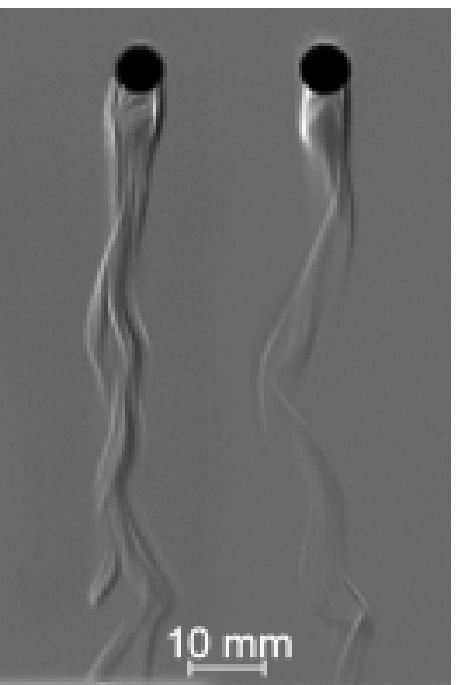}\\
\footnotesize{(c)}
\end{center}
\end{minipage}
\hfill
\begin{minipage}[t]{3.0cm}
\begin{center}
\includegraphics[width=3.0cm]{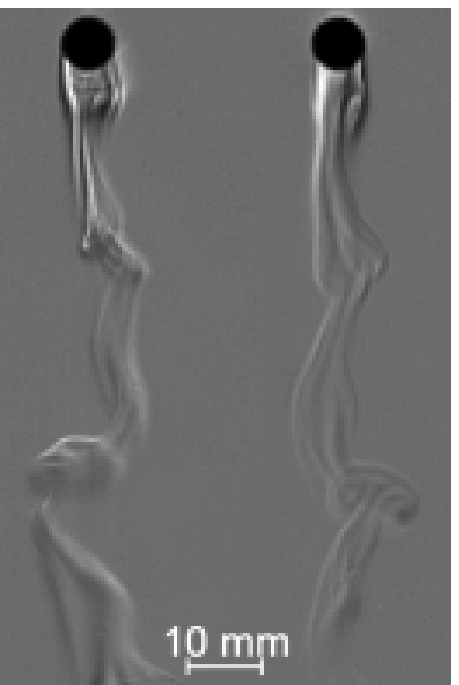}\\
\footnotesize{(d)}
\end{center}
\end{minipage}\\
\caption{Stereoscopic views of falling and rising spheres and their
wakes. The left part of each frame shows the xz-plane and the right
part the yz-plane. In each case the sphere diameter is 6 mm. The
values of the parameters $\rho_s/\langle \rho \rangle$, $G$, and
$Re_T$ are, respectively: ({\em a}) 1.04, 359, 546; ({\em b}) 1.05,
394, 608; ({\em c}) 0.96, 355, 647; ({\em d}) 0.95, 390, 656.}
\label{FallingRisingFig2}
\end{figure}

Figure \ref{FallingRisingFig2} shows stereoscopic images of the wake
structure behind falling spheres with densities approximately 4\%
(fig.\ \ref{FallingRisingFig2}{\em a}) and 5\% (fig.\
\ref{FallingRisingFig2}{\em b}) higher than that of the surrounding
liquid, and, for comparison, that behind rising spheres with
densities that are approximately 4\% (fig.
\ref{FallingRisingFig2}{\em c}) and 5\% (fig.
\ref{FallingRisingFig2}{\em d}) lower. In all these examples the
sphere diameter is 6 mm, so that the parameter $G$ is roughly
identical in cases ({\em a}) and ({\em c}), and in cases ({\em b})
and ({\em d}). The lighter spheres have a slightly higher vertical
velocity than the heavier spheres, as indicated by $Re_T$ in table
\ref{Table1}. The wakes of the falling spheres appear to have a more
`irregular' structure, and the path followed by these spheres shows
much larger deviations from a straight vertical line. These pictures
illustrate that the density ratio $\rho_s/\langle \rho \rangle$
matters, even at values close to 1.

\begin{figure}[p]
\unitlength1cm
\begin{minipage}[t]{4.0cm}
\begin{center}
\includegraphics[width=3.68cm]{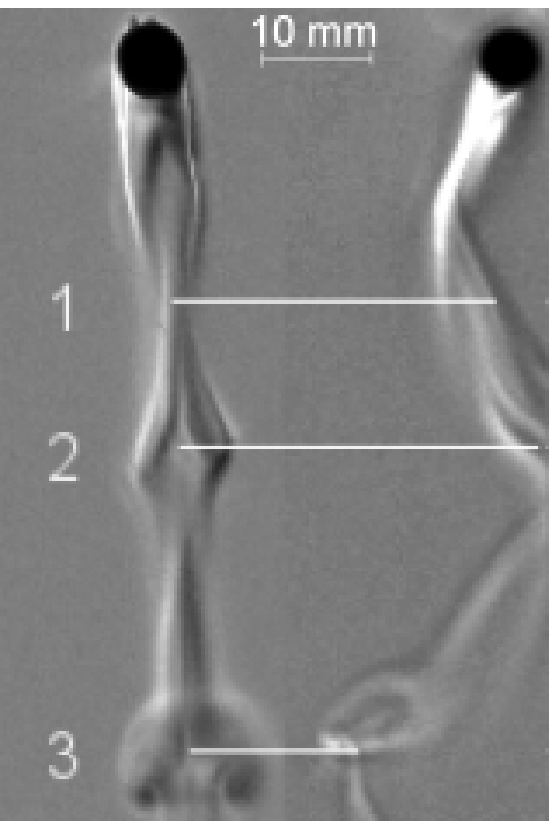}
\footnotesize{(a)}
\end{center}
\end{minipage}
\hfill
\begin{minipage}[t]{3.5cm}
\begin{center}
\includegraphics[width=3cm]{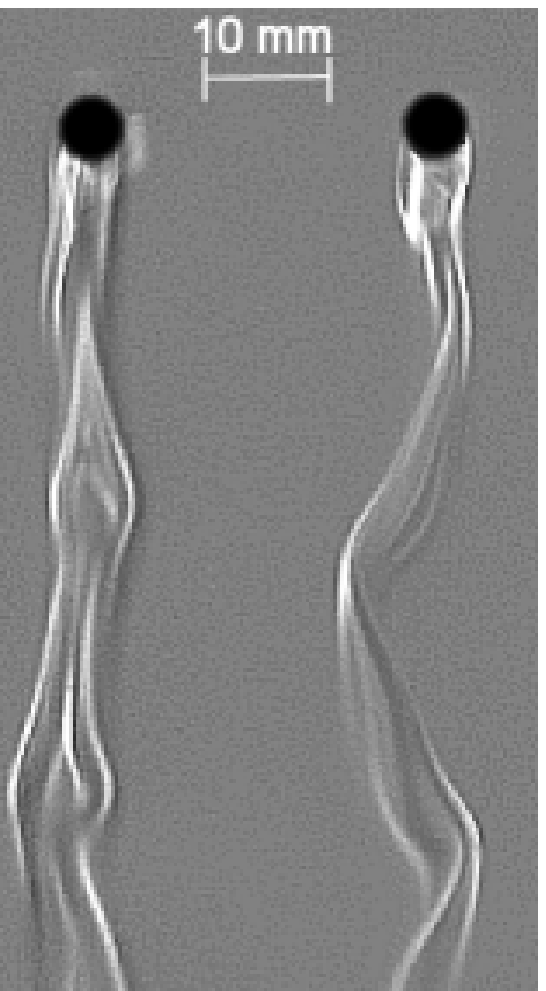}
\footnotesize{(b)}
\end{center}
\end{minipage}
\hfill
\begin{minipage}[t]{4cm}
\begin{center}
\includegraphics[width=3.3cm]{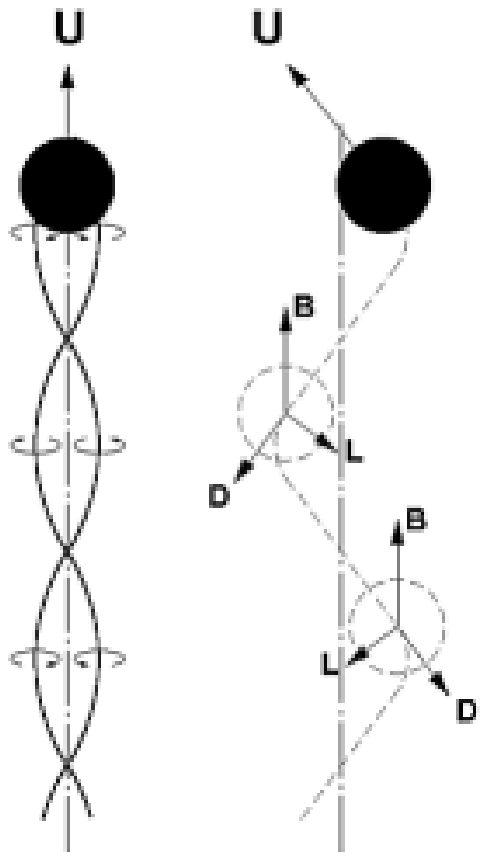}\\
\footnotesize{(c)}
\end{center}
\end{minipage}\\
\caption{Stereoscopic views of rising zigzagging spheres and their
wakes. The left part of each frame shows the xz-plane and the right
part the yz-plane. The views illustrate the crossing at the
center-line of the zigzag path of the two counter-rotating threads
of the wake (`1'), the occurrence of kinks (`2') at the extremes of
the path, and the formation of hairpin-like vortices (`3') as two
neighbouring kinks connect. Values of the parameters $d$,
$\rho_s/\langle \rho \rangle$, $G$ and $Re_T$ are, respectively:
({\em a}) 5 mm, 0.95, 297, 450; ({\em b}) 5 mm, 0.95, 306, 475. As
shown in ({\em c}) the crossing of the vortex threads results in a
lift force $L$ that is always directed towards the center-line of
the zigzag path. D shows the direction of the drag force and B the
one of the buoyancy.} \label{RisingFig3} \label{RisingCrossingFig3}
\end{figure}

Figures \ref{RisingCrossingFig3}{\em a} and
\ref{RisingCrossingFig3}{\em b} give examples of a phenomenon that
we believe to be characteristic for spheres following a zigzag path,
namely that the two counter-rotating threads in the wake cross at
the centerline of the zigzag (indicated by `1' in fig.
\ref{RisingCrossingFig3}{\em a}). The presence of these threads of
opposite-signed streamwise vorticity implies that the sphere
experiences a lift force. As a consequence of the periodic crossing
of the threads this force is always directed towards the zigzag
center-line (see the sketch in fig.\ref{RisingCrossingFig3}{\em c}).
A similar observation was made by De Vries \emph{et al.}
\cite{Vries} on the wake behind zigzagging gas bubbles.

\begin{center}
\begin{figure}[p]
\begin{center}
\includegraphics[angle=0, width=0.8\textwidth]{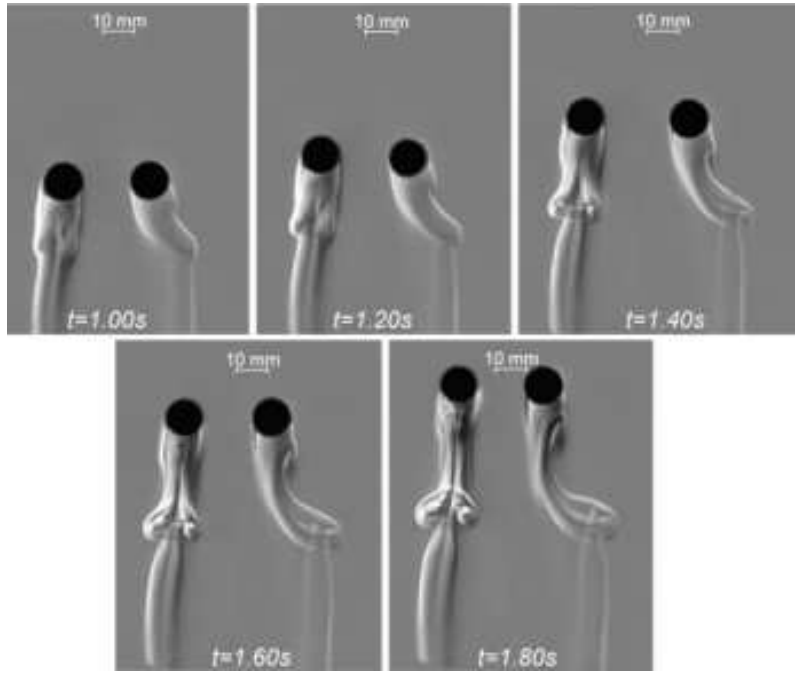}\\
\caption{Sequence of stereoscopic views of a rising sphere and its
wake. The left part of each frame shows the xz-plane and the right
part the yz-plane. The views illustrate the process of formation of
a hairpin-like vortex ($d =10$ mm, $\rho_s/\langle \rho \rangle=
0.99$, $G=350$ and $Re_T=576$).} \label{RisingKinkFig4}
\end{center}
\end{figure}
\end{center}

Schouveiler \& Provansal \cite{Schouveiler} remark that for a fixed
sphere ``the dynamics of the two opposite-sign streamwise vortices
... presents a striking similarity with the long-wavelength (or
Crow) instability of a pair of counter-rotating parallel vortices"
and further ``such a vortex pair instability could be responsible of
the appearance of unsteadiness in the sphere wake". Figure
\ref{RisingCrossingFig3} suggests that the situation is slightly
different for freely moving spheres. Here it appears that close to
the sphere each of the vortices first develop a `kink' (indicated by
`2' in fig. \ref{RisingCrossingFig3}{\em a}), a process in which the
curvature of the vortices presumably plays an important role
\cite{Betchov}. As the kinks develop further downstream of the
sphere they come near each other and finally combine into what
resembles a hairpin vortex (indicated by `3'). This sequence of
events can be seen in the flow visualizations presented in figure
\ref{RisingKinkFig4}, see also figure~6 of ref. \cite{Magarvey1}.

As the kinks develop and hairpin-like vortices are formed further
downstream, a pattern results. Lunde \& Perkins \cite{Lunde}
interpreted this pattern as a series of hairpin vortices of
alternating sign, shed periodically by the spheres at the extremes
of the zigzag path. Our visualizations suggest instead that the
streamwise vorticity produced at the surface of the sphere does not
change sign; the legs of the like-signed hairpin vortices cross at
the centerline of the zigzag.

\begin{figure}[!h]
\begin{center}
\includegraphics[angle=0, width=1.0\textwidth]{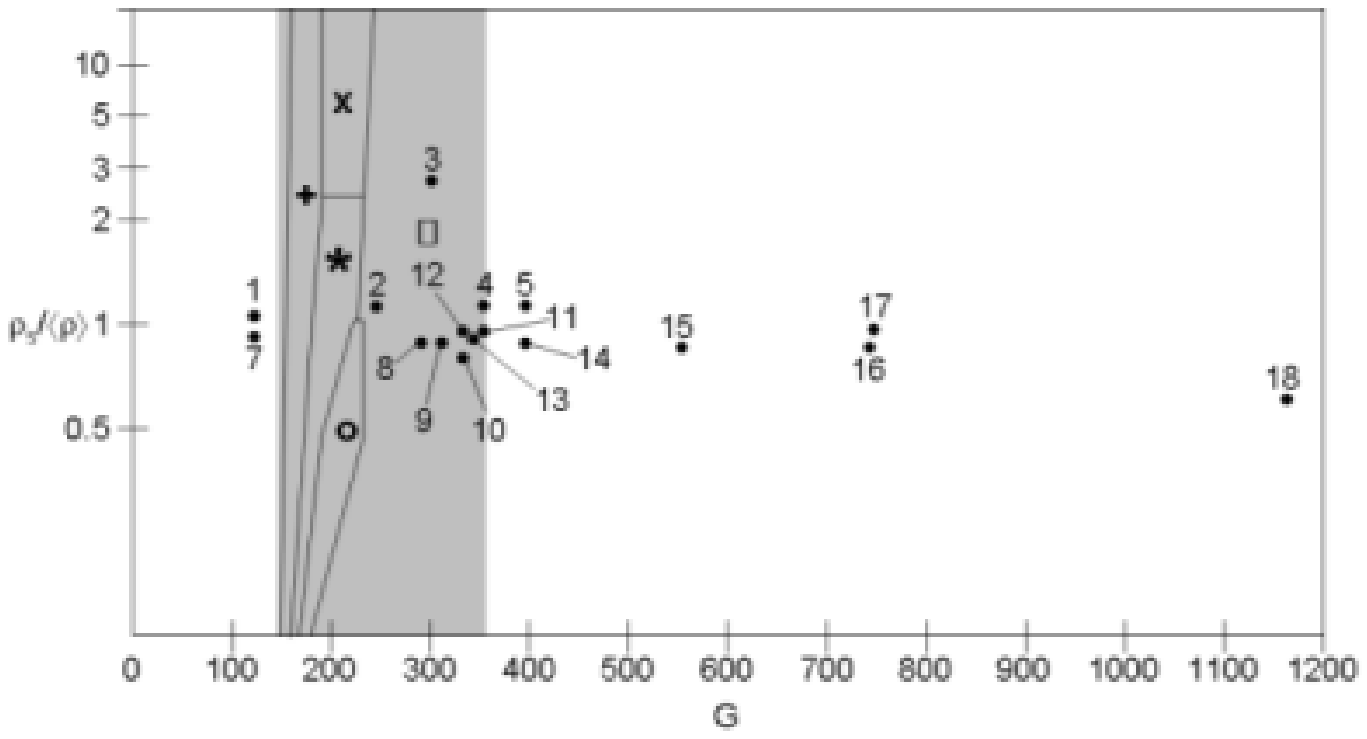}\\
\caption{Phase diagram: density ratio $\rho_s/\langle \rho \rangle$
versus Galileo number. The grey box is the regime analysed by Jenny
\emph{et al.} \cite{Jenny}. They find in the most left region an
axisymmetric wake. The symbols, directly taken from \cite{Jenny},
denote: $+$ steady and oblique, $*$ oblique and oscillating regime
with low frequency ($0.045 \leq f \leq 0.068$), $\times$ oblique and
oscillating regime with high frequencies ($f = 0.180$), O zigzagging
periodic regime ($0.023 \leq f \leq 0.035$) and $\square$ chaotic
regime. The numbers denote the number of our experiment in table 1.
Experiments 6
and 19  fall outside the diagram.} \label{DusekDiagram}
\end{center}
\end{figure}

Figure \ref{FallingRisingFig2}{\em b} is an example in which more
than one kink develops in a half-period of the zigzag. We have not
yet been able to determine the conditions (in terms of the
parameters $\rho_s/\langle \rho \rangle$, $G$, or $Re_T$) that
select the number of kinks that are formed. What is remarkable is
that the development of the kinks and the subsequent formation of
the hairpin vortices do not seem to affect the trajectory of the
sphere. This corroborates the opinion that at high Reynolds numbers
the details of the vorticity distribution very close to a body
basically determine the forces that it experiences.
We will now turn to experiments with density ratios more different
from one. Recently Jenny \emph{et al.} \cite{Jenny1} reported on
their numerical work on freely moving spheres in a Newtonian fluid.
They focussed on the frequencies in the wake and the path of the
sphere in the parameter space spanned up by the density ratio and
the Galileo number. Figure \ref{DusekDiagram} reproduces their phase
diagram. The numbers in the diagram refer to the numbers of the
experiments given in table \ref{Table1}. A lot of our experiments
are outside their investigated region and new experiments should be
done in the interesting regions around a Galileo number of 200.
Further experiments should focus more on the frequencies in the wake
of the sphere and compare this to the frequencies given by Jenny
\emph{et al.} (see caption of figure \ref{DusekDiagram}).
Furthermore, we must stress that wake visualizations with the
Schlieren method demand a temperature gradient in the water. Hence
the density and viscosity of the water are not constant through the
entire flow field and the local Galileo number will not be constant.
The differences between the \emph{mean} Galileo number and the local
Galileo number can reach 10 \% and must be taken into account
when analyzing figure \ref{DusekDiagram}.

A striking difference between our experimental data and the
numerical data of Jenny \emph{et al.} is the behavior of falling
spheres with a density ratio close to one. From figures
\ref{FallingRisingFig2} and \ref{ExtraPicturesFalling} it can be
seen that these falling spheres can also fall in a non-vertical
path. This contradicts Jenny \emph{et al.} who claim that only
rising spheres can go in a zigzagging motion (the circles in the
phase diagram figure \ref{DusekDiagram}).

\begin{figure}[p]
\unitlength1cm
\begin{minipage}[t]{2.6cm}
\begin{center}
\includegraphics[width=2.6cm]{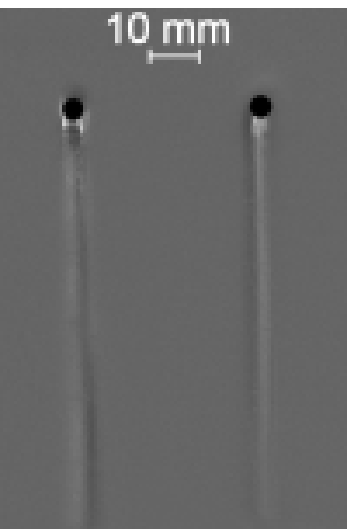}\\
\footnotesize{(a)}
\end{center}
\end{minipage}
\hfill
\begin{minipage}[t]{2.6cm}
\begin{center}
\includegraphics[width=2.6cm]{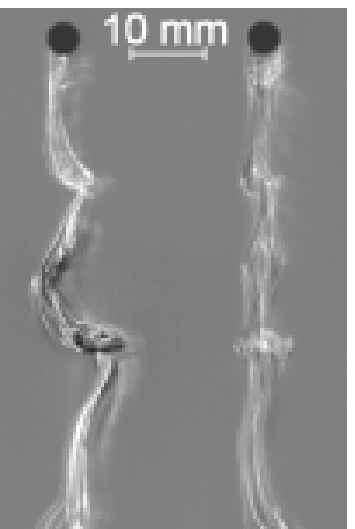}\\
\footnotesize{(b)}
\end{center}
\end{minipage}
\hfill
\begin{minipage}[t]{2.6cm}
\begin{center}
\includegraphics[width=2.6cm]{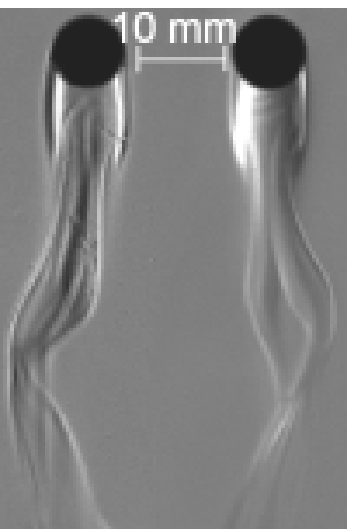}\\
\footnotesize{(c)}
\end{center}
\end{minipage}
\hfill
\begin{minipage}[t]{2.6cm}
\begin{center}
\includegraphics[width=2.6cm]{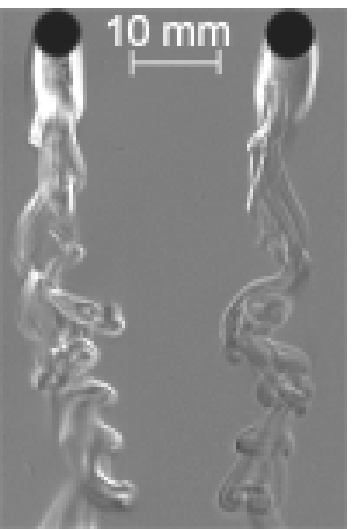}\\
\footnotesize{(d)}
\end{center}
\end{minipage}
\caption{Stereoscopic views of rising spheres and their wake
structures observed at several Reynolds numbers (Reynolds number
increases from a to d and continues in figure
\ref{ExtraPicturesRising2}). The left part of each frame shows the
xz-plane and the right part the yz-plane. Values of the parameters
$d$, $\rho_s/\langle \rho \rangle$, $G$ and $Re_T$ are,
respectively: ({\em a}) 3.2 mm, 0.97, 121, 210; ({\em b}) 4.0 mm,
0.88, 334, 565; ({\em c}) 8.0 mm, 0.99, 331, 602; ({\em d}) 6.4 mm,
0.93, 534, 920.} \label{ExtraPicturesRising1}
\end{figure}

\begin{figure}[p]
\unitlength1cm
\begin{minipage}[t]{2.6cm}
\begin{center}
\includegraphics[width=2.6cm]{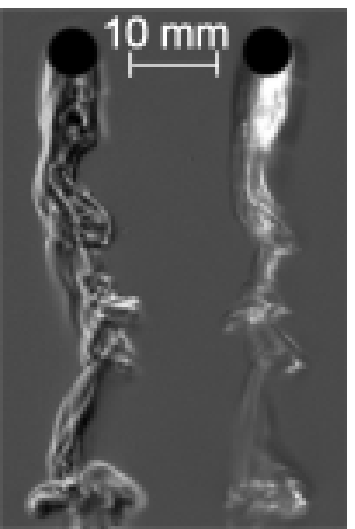}\\
\footnotesize{(a)}
\end{center}
\end{minipage}
\hfill
\begin{minipage}[t]{2.6cm}
\begin{center}
\includegraphics[width=2.6cm]{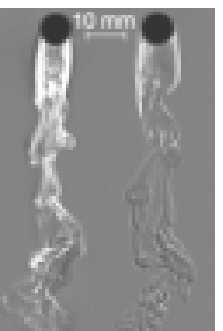}\\
\footnotesize{(b)}
\end{center}
\end{minipage}
\hfill
\begin{minipage}[t]{2.6cm}
\begin{center}
\includegraphics[width=2.6cm]{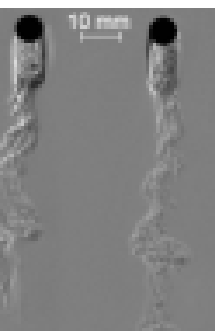}\\
\footnotesize{(c)}
\end{center}
\end{minipage}
\hfill
\begin{minipage}[t]{2.6cm}
\begin{center}
\includegraphics[width=2.6cm]{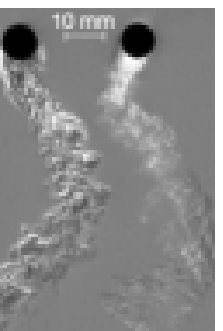}\\
\footnotesize{(d)}
\end{center}
\end{minipage}
\caption{Stereoscopic views of rising spheres and their wake
structures observed at several Reynolds numbers (Reynolds number
increases from a to d). The left part of each frame shows the
xz-plane and the right part the yz-plane. Values of the parameters
$d$, $\rho_s/\langle \rho \rangle$, $G$ and $Re_T$ are,
respectively: ({\em a}) 6.4mm, 0.87, 728, 1180; ({\em b}) 7.9 mm,
0.93, 732, 1350; ({\em c}) 6.4 mm, 0.65, 1160, 1965; ({\em d}) 9.5
mm, 0.50, 2548, 4623.} \label{ExtraPicturesRising2}
\end{figure}

\begin{figure}[p]
\unitlength1cm
\begin{minipage}[t]{2.6cm}
\begin{center}
\includegraphics[width=2.6cm]{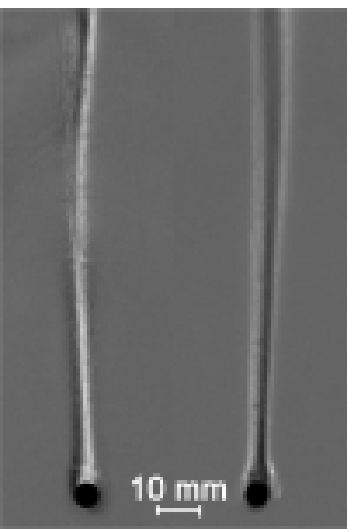}\\
\footnotesize{(a)}
\end{center}
\end{minipage}
\hfill
\begin{minipage}[t]{2.6cm}
\begin{center}
\includegraphics[width=2.6cm]{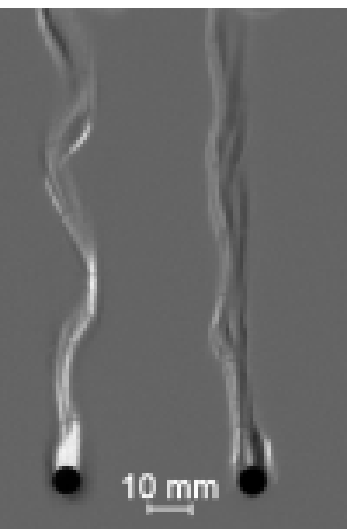}\\
\footnotesize{(b)}
\end{center}
\end{minipage}
\hfill
\begin{minipage}[t]{2.6cm}
\begin{center}
\includegraphics[width=2.6cm]{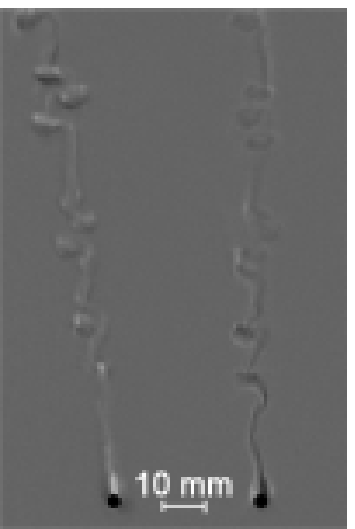}\\
\footnotesize{(c)}
\end{center}
\end{minipage}
\hfill
\begin{minipage}[t]{2.6cm}
\begin{center}
\includegraphics[width=2.6cm]{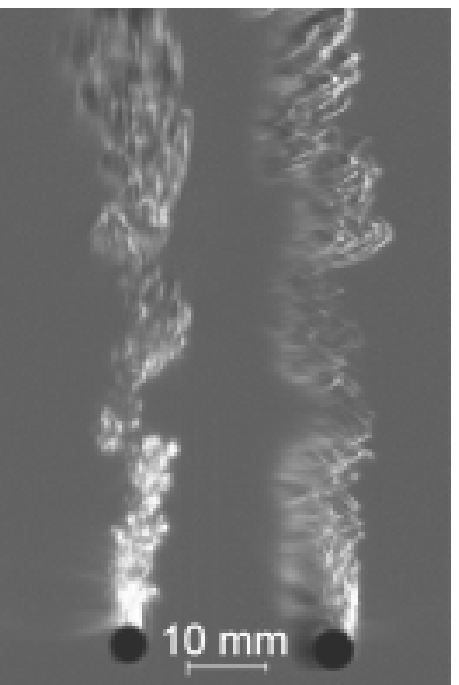}\\
\footnotesize{(d)}
\end{center}
\end{minipage}
\caption{Stereoscopic views of falling spheres and their wake
structures observed at several Reynolds numbers (Reynolds number
increases from a to d). The left part of each frame shows the
xz-plane and the right part the yz-plane. Values of the parameters
$d$, $\rho_s/\langle \rho \rangle$, $G$ and $Re_T$ are,
respectively: ({\em a}) 3.2mm, 1.03, 121, 205; ({\em b}) 4.0 mm,
1.06, 239, 325; ({\em c}) 1.5 mm, 2.79, 304, 450; ({\em d}) 4.0 mm,
2.63, 1261, 1970.} \label{ExtraPicturesFalling}
\end{figure}

From our experiments one concludes that for increasing Reynolds
number the wake becomes more irregular (figures
\ref{ExtraPicturesRising1} to \ref{ExtraPicturesFalling}). The
two-threaded wake structure is also present for higher Reynolds
numbers. Is the double threaded wake structure also present in the
case of the highest Reynolds numbers, where the wake structure has a
turbulent structure? If so, do instabilities in the wake cause
kinking of the vortex threads which leads to this turbulent wake
structure? Further research will address these questions in order to
get a better understanding of the boundary layer separation from
spheres at high Reynolds numbers as shown on this year's cover.

\section{Conclusions}
Flow visualizations of the wakes behind solid spheres moving under
the action of gravity reveal remarkable differences with the wakes
behind spheres held fixed: the crossing of threads of
opposite-signed vorticity, the formation of kinks on these threads
that develop into hairpin vortices. The ratio between the densities
of the sphere and that of the surrounding fluid appears to be
important. Our experiments clearly show the difference in path and
wake structure between rising and falling spheres with the same
Galileo number. Furthermore, the double threaded wake structure
seems to be a basic feature, even for large Reynolds numbers. This
should be investigated thoroughly in future research.

\section{Acknowlegdments}
The authors
thank
Andrea Prosperetti,
M. Versluis, and C.D. Ohl for
helpful discussions and suggestions and G.-W. Bruggert and H.
Scholten for the technical support.
This work is part of the research programme of the
Stichting voor Fundamenteel Onderzoek der Materie (FOM), which is
financially supported by the Nederlandse Organisatie voor
Wetenschappelijk Onderzoek (NWO).
%\newpage

\end{document}